\documentstyle[preprint,aps]{revtex}
\begin{document}
\draft
\tighten
\title{Chemical  equilibration  in  viscous  quark-gluon plasma and
electromagnetic signals}
\author{\bf A. K. Chaudhuri}
\address{ Variable Energy Cyclotron Centre\\
1/AF,Bidhan Nagar, Calcutta - 700 064\\}
\maketitle
\begin{abstract}
We   investigate   the   chemical  equilibration  of  the  parton
distributions in collisions of two  heavy  nuclei,  assuming  the
partonic  fluid  to  be  ideal  as  well  as viscous. The initial
conditions  are  taken  from  HIJING   calculations   for   Au+Au
collisions  at  RHIC  and LHC energies. It was seen that when the
viscous drag is taken into account in the fluid  flow,  the  life
time  of  the  plasma  is  increased by nearly a factor of 2. The
temperature as well as fugacities evolve slowly than their  ideal
counterpart.  The  photon  and  lepton  pair  production was also
investigated. There is a two fold  increase  in  the  photon  and
lepton  pair numbers with viscosity on. The increase in the large
$p_T$ photons and the large invariant mass lepton pairs  are  due
to slower rate of temperature evolution.
\end{abstract}


\section{Introduction}

In  relativistic  heavy  ion  collisions,  a strongly interacting
matter called quark-gluon plasma (QGP) is expected to be  formed.
The  two  colliding  nuclei  can  be  visualized as two clouds of
valence and sea  partons,  which  pass  through  each  other  and
interact  \cite{ge92}.  The  multiple  parton collisions can then
produce a dense plasma of quarks and gluons. After its formation,
the plasma will expand, cool and become more dilute.  If  quantum
chromodynamics  admits  first order deconfinement or chiral phase
transition, it is likely that the  system  will  pass  through  a
mixed  phase  of  quarks,  gluons and hadrons, before the hadrons
lose thermal contact at sufficient  dilution  and  stream  freely
towards the detectors.

Several  questions  about  the  structure of the matter formed in
these nuclear collisions arise. Does the initial partonic  system
attain  kinetic  equilibrium? Probably yes, as the initial parton
density is large and the partons suffer many collisions in a very
short time \cite{es94}. Does it attain chemical equilibrium? This
will       depend       on       the        time        available
\cite{bi93,ge93,ge93b,al94,wo96} to the partonic system before it
convert  into  a  mixed  phase  or  before perturbative QCD is no
longer  applicable.  The  time  available  for  equilibration  is
perhaps too short (3-5 fm/c) at the energies ($\sqrt{s} \leq$ 100
GeV) accessible to the Relativistic Heavy Ion Collider (RHIC). At
the  energies  ($\sqrt{s}  \leq$  3  TeV/nucleon)  that  will  be
achieved at CERN Large Hadron Collider(LHC), this time  could  be
large  (more  than  10 fm/c). If one consider only a longitudinal
expansion of the system, the QGP formed  at  LHC  energies  could
approach chemical equilibrium very closely, due to higher initial
temperature  predicted  to  be  attained  there.  The question of
chemical  equilibrium   has   the   added   importance   as   the
preequilibrium   phase   may   influence  the  yield  of  certain
quark-gluon plasma signals, such as  lepton  pairs,  photons  and
hadron containing heavy quarks.

Space-time  evolution  of  quark and gluon distributions has been
investigated  in  the   framework   of   parton   cascade   model
\cite{ge92},  which  is  based  on  the concept of inside-outside
cascade \cite{an80,hw86,bl87} and evolve parton distributions  by
Monte-Carlo  simulation  of  a  relativistic  transport  equation
involving lowest order perturbative  QCD  scattering  and  parton
fragmentations.  Early  calculations  were done by assuming fixed
$p_{T}$ and virtuality cutoffs for the partonic  interactions  to
ensure  the  applicability  of  the perturbative expansion of QCD
scattering      process.       From       numerical       studies
\cite{ge93,ge92b,ka92,mu93,sh92,mu92}  three  distinct  phases of
parton evolution can be distinguished: (1) Gluon thermalise  very
rapidly,   reaching   approximately   isotropic   momentum  space
distribution after a time of the order  of  0.3  fm/c.  (2)  Full
equilibration  of  gluon  phase  space density takes considerably
longer. (3) The evolution of quark distributions lags behind that
of the gluons, the relevant QCD cross sections  being  suppressed
by  a  factor  of  2-3.  In the recently formulated self-screened
parton cascade model \cite{es96} early hard scattering produces a
medium which screens the longer ranged  color  fields  associated
with  softer  interactions.  When  two  heavy  nuclei  collide at
sufficiently high energy, the screening occurs at a length  scale
where  perturbative  QCD  still  applies.  This  approach  yields
predictions for initial conditions of the  forming  QGP,  without
the  need  for  any  ad-hoc  momentum and virtuality cutoff. This
calculations also indicate that the QGP likely to  be  formed  in
such collisions are far from chemical equilibrium.

Essentially  same  picture  emerges  from the study of Biro et al
\cite{bi93}. Their calculations were motivated to obtain a  lucid
understanding  of  the  dependence  of  different  time scales on
various parameters and model assumptions and  a  better  physical
understanding  of  the  infrared  cut  off required in the parton
cascade model .The parton cascade model is a complex model and it
is difficult to unravel these dependence from model calculations.
They assumed that after a time  $\tau_{iso}$,  the  evolution  of
partonic  system  is governed by the energy-momentum conservation
law. They derived a set of rate equations describing the chemical
equilibration of gluons and quarks, including the  medium  effect
on relevant transport coefficients. The initial conditions of the
partonic  system  was  obtained  from  HIJING calculations. Their
results support the picture emerging from parton  cascade  model,
that  the  plasma  is  essentially  a  gluon  plasma.  The  gluon
distributions attain near chemical equilibrium at  LHC  energies,
and  the  quark  distributions  lag behind. At RHIC energies, the
gluon as well as quark distributions are far from equilibrium. In
their calculation  the  partonic  fluid  is  assumed  to  undergo
longitudinal  expansion  only.  Effect of transverse expansion on
chemical equilibration was studied in ref.[\cite{sr97,sr97b}]. At
RHIC  energies,  results  for  pure  longitudinal  expansion  and
transverse  expansion are quite similar except at large values of
transverse radius.  However,  at  LHC  energies,  the  transverse
expansion changes the scenario, the plasma initially approach the
chemical  equilibrium,  but then is driven away from it, when the
transverse  velocity  gradients  develop.  Interestingly,   total
parton multiplicity as well as lepton pair and photon productions
are not much affected by the transverse expansion.

In    all    these   calculations   \cite{bi93,sr97,sr97b},   the
hydrodynamic evolution of the partonic system was assumed  to  be
ideal.  It  is well known that dissipative effects like viscosity
can affect the flow \cite{da85,ho85,ch95} . The system will  cool
slowly as viscous drag will oppose the expansion. As the chemical
equilibration  rate  depend  on  the  cooling  rate, it will also
change. In the present paper,  we  will  explore  the  effect  of
viscosity   on   the   evolution   and  equilibration  of  parton
distributions. We will also investigate its effect on the  lepton
pair and photon production, typical QGP signals.

The  paper  is  organised as follows: in section 2, we set up the
basic equations for hydrodynamic and chemical  evolution  of  the
plasma, also define the viscosity coefficients. In section 3, the
results  will  be discussed. In section 4, we study the effect of
viscosity on photon and dilepton  productions.  Summary  will  be
given in section 5.

\section{Hydrodynamic   expansion   and  chemical  equilibration}
\subsection{Basic Equations}

We  assume  that  after the collision, a {\em symmetric} partonic
system (QGP) is formed in the central rapidity  region.  We  also
assume that the partonic system achieves a kinetic equilibrium by
time  $\tau_{iso}$,  when  the  momenta of partons became locally
isotropic \cite{bi93}. Local isotropy in momentum can be said  to
be   established  when  variance  in  the  longitudinal  momentum
distribution equals  the  variance  in  the  transverse  momentum
distribution.   This   occurs   after  a  proper  time  of  about
$0.7\lambda_f$ \cite{es94}, $\lambda_f$ being the  partonic  mean
free  path.  At  collider  energies,  this  time  corresponds  to
$\tau_{iso}\approx 0.2-0.3$ fm/c \cite{bi93}. Beyond $\tau_{iso}$
further expansion of the partonic  system  can  be  described  by
hydrodynamic equations. The approach to chemical equilibration is
then governed by a set of master equation which are driven by the
two    body    reactions($gg\leftrightarrow    qq$)   and   gluon
multiplication   and   its   inverse   process,   gluon    fusion
($gg\leftrightarrow  ggg$). The hot matter continue to expand and
cool due  to  expansion  and  chemical  equilibration,  till  the
temperature  falls below the critical value ($T_c=160$ MeV), when
we terminate the evolution.

We assume that the hydrodynamic expansion is purely longitudinal.
As indicated in the analysis \cite{sr97,sr97b}, at RHIC energies,
the  transverse  expansion  effect is minimal. It does affect the
parton equilibration rate at LHC  energies,  the  effect  showing
sensitive  dependence on the initial condition of the plasma. For
the initial condition as obtained from HIJING calculations,  this
effect is not large \cite{sr97,sr97b}.

For   boost-invariant   longitudinal  flow,  the  energy-momentum
conservation  equation  for  the  partonic  fluid  including  the
viscosity, is the well known equation \cite{da85,ho85,ch95},

\begin{equation}
\frac{d\varepsilon}{d\tau}+\frac{\varepsilon+P-\frac{4/3\eta+\zeta}
{\tau}}{\tau}=0 \label{1}
\end{equation}

\noindent  where  $\eta$  and  $\zeta$  are  the  shear  and bulk
viscosity  coefficients.  In  the   QGP,   the   bulk   viscosity
coefficient is zero. It is apparent that in one dimensional flow,
the  effect of viscosity is to reduce the pressure. One must note
that, under no circumstances, the viscous  drag  can  exceed  the
pressure.

Equation   of   state  need  to  be  specified  for  solving  the
hydrodynamic equation \ref{1}. For partially equilibrated  plasma
of  massless  particles,  the equation of state can be written as
\cite{bi93}

\begin{equation}   \varepsilon=3P=[a_2\lambda_g  +  b_2(\lambda_q
+\lambda_{\bar{q}})]T^4 \label{2} \end{equation}

\noindent  which  imply  a  speed  of  sound $c_s=1/\sqrt{3}$. In
Eq.\ref{2}, $a_2=8\pi^2/15$, $b_2=7\pi^2N_f/40$, $N_f\approx 2.5$
is the number of  dynamical  quark  flavor.  $\lambda_i$  is  the
fugacity  for  the  parton  species $i$. Here we have defined the
fugacities through the relations:

\begin{mathletters}     \begin{eqnarray}    n_g    =    \lambda_g
\tilde{n}_g\\  n_q  =  \lambda_q  \tilde{n}_q\\   n_{\bar{q}}   =
\lambda_{\bar{q}}  \tilde{n}_{\bar{q}}  \end{eqnarray}  \label{3}
\end{mathletters}

\noindent  where $\tilde{n}_i$ is the equilibrium density for the
parton species $i$:

\begin{mathletters}                              \begin{eqnarray}
\tilde{n}_g&&=\frac{16}{\pi^2}\zeta(3)T^3=a_1T^3\\
\tilde{n}_q&&=\frac{9}{2\pi^2}\zeta(3)N_fT^3=b_1T^3
\end{eqnarray} \label{4} \end{mathletters}

We  further  assume  that $\lambda_q=\lambda_{\bar{q}}$. Denoting
the derivative with respect to $\tau$ by an  overdot,  eq.\ref{1}
can be rewritten as,

\begin{equation}    \frac   {\dot{\lambda_g}   +b(\dot{\lambda_q}
+\dot{\lambda_{\bar{q}}})}        {\lambda_g         +b(\lambda_q
+\lambda_{\bar{q}})}+     4\frac{\dot{T}}{T}     +\frac{4}{3\tau}
-\frac{4}{3}\frac{\eta}{\tau^2}                     \frac{1}{[a_2
\lambda_g+b_2(\lambda_q+\lambda_{\bar{q}})]T^4}    =0   \label{5}
\end{equation}

\noindent  where  $b=b_2/a_2=21N_f/64$.  If the partonic fluid is
considered to be ideal fluid, $\eta=0$ and the above equation can
be integrated to obtain,

\begin{equation}         [\lambda_g         +         b(\lambda_q
+\lambda_{\bar{q}})]^{3/4}     T^3\tau     =const.      \label{6}
\end{equation}

For          a         fully         equilibrated         plasma,
($\lambda_g=\lambda_q=\lambda_{\bar{q}}=1$), and  we  obtain  the
Bjorken   scaling   law   $T^3\tau=const.$   For  viscous  fluid,
eq.\ref{5}  can  not  be  integrated  and  has  to  be  evaluated
numerically.

The   master   equation   for   the  dominant  reaction  channels
$gg\leftrightarrow  ggg$  and  $gg\leftrightarrow  q\bar{q}$  for
chemical equilibration are,

\begin{eqnarray} \partial_\mu (n_g u^\mu)=&&n_g (R_{2 \rightarrow
3}-R_{3  \rightarrow  2})-  (n_g  R_{g  \rightarrow q} - n_q R_{q
\rightarrow  g})\\  \partial_\mu  (n_q  u^\mu)  =&&  \partial_\mu
(n_{\bar{q}}  u^\mu)  =  (n_g  R_{g  \rightarrow  q}  -  n_q R_{q
\rightarrow g}) \label{7} \end{eqnarray}

\noindent  in an obvious notation. Using Juttner distribution for
the phase space distribution for the partons, the  rates  can  be
factorized as \cite{bi93},

\begin{mathletters}      \begin{eqnarray}     n_g(R_{2\rightarrow
3}-R_{3\rightarrow        2})&&=\frac{1}{2}\sigma_3         n_g^2
(1-\frac{n_g}{\tilde{n}_g})\\   n_g   R_{g\rightarrow   q}-   n_q
R_{q\rightarrow  g}&&=\frac{1}{2}  \sigma_2  n_g^2   (1-\frac{n_q
n_{\bar{q}}   \tilde{n}_g^2}   {\tilde{n}_g   \tilde{n}_{\bar{q}}
n_g^2}) \end{eqnarray} \label{8} \end{mathletters}

\noindent where $\sigma_3$ and $\sigma_2$ are thermally averaged,
velocity weighted cross sections,

\begin{mathletters}       \begin{eqnarray}      \sigma_3      &&=
<\sigma(gg\rightarrow ggg)v>\\ \sigma_2 &&= <\sigma(gg\rightarrow
q\bar{q})v>, \end{eqnarray} \label{9} \end{mathletters}

Defining       density       weighted       reaction       rates;
$R_3=\frac{1}{2}\sigma_3 n_g$ and $R_2=\frac{1}{2}\sigma_2  n_g$,
eq.\ref{7} can be written as,

\begin{eqnarray}    \frac{\dot{\lambda}_g}    {\lambda_g}   +   3
\frac{\dot{T}}{T}+\frac{1}{\tau} &&= R_3 (1-\lambda_g)  -  2  R_2
(1-\frac{\lambda_q       \lambda_{\bar{q}}}      {\lambda_g^2})
\label{10a}\\
\frac{\dot{\lambda}_q}          {\lambda_q}          +          3
\frac{\dot{T}}{T}+\frac{1}{\tau}    &&=    R_2    \frac{a_1}{b_1}
(\frac{\lambda_g}{\lambda_q}-\frac{\lambda_{\bar{q}}}
{\lambda_g}) \label{10b} \end{eqnarray}

Biro  et  al \cite{bi93} have evaluated the reaction rates $R_2$
and $R_3$, taking into account the color Debye screening and  the
Landau-Pomeranchuk-Migdal  effect  suppressing  the induced gluon
radiation. The results are,

\begin{eqnarray}  R_2  &&\approx  0.24  N_f \alpha_s^2\lambda_g T
\ln(1.65/\alpha_s\lambda_g) \label{11a}\\
R_3
&&=2.1\alpha_s^2T(2\lambda_g-\lambda_g^2)^{1/2}        \label{11b}
\end{eqnarray}

The   coupled   Eqs(\ref{5},\ref{10a},\ref{10b})   determine  the
evolution of $T(\tau)$, $\lambda_g(\tau)$  and  $\lambda_q(\tau)$
towards  the  chemical equilibrium and are solved numerically. It
may be noted that the rate  equations  (\ref{10a},\ref{10b})  may
give  negative  growth  rate  for the fugacities, in case of very
slow cooling. For  viscous  fluid,  viscosity  being  temperature
dependent,  the  cooling is slowed down, and such a situation may
arise at early times when the temperature of the partonic  system
is  very  large.  However,  physics  of  the problem require that
fugacities    can    grow    only.    Thus,     while     solving
eqs(\ref{10a},\ref{10b}),  we  impose  the  additional constraint
that,

\begin{equation}    \dot{\lambda}_{g,q}    \geq    0   \label{12}
\end{equation}

\subsection{Viscosity coefficient}

Several  results  for  viscosity  coefficients  for QGP system is
available \cite{da85,ho85,ga85,ba90,th91}. Thoma \cite{th91}  has
calculated  the  shear  viscosity  coefficient of the QGP, in the
relaxation time approximation. Screening effects were taken  into
account  by  using an effective perturbation theory developed for
the  finite  temperature  QCD,  in  the   weak   coupling   limit
\cite{br90}.   His   results  agree  well  with  the  variational
calculation of Baym et al \cite{ba90}, and will be  used  in  the
present calculation.

\begin{mathletters}          \begin{eqnarray}         \eta_q=0.82
\frac{T^3}{\alpha_x^2       \ln(1/\alpha_s)}\\        \eta_g=0.20
\frac{T^3}{\alpha_x^2  \ln(1/\alpha_s)} \end{eqnarray} \label{13}
\end{mathletters}

For   plasma,   away   from   equilibrium,  the  shear  viscosity
coefficient for QGP then can be written as,

\begin{equation}    \eta=\lambda_g   \eta_g   +\lambda_q   \eta_q
\label{14} \end{equation}

\noindent with $\eta_{g,q}$ given in eq.\ref{13}.

We  have assumed that the dominant reactions towards the chemical
equilibrium    are    the     $gg\leftrightarrow     ggg$     and
$gg\leftrightarrow  q\bar{q}$  reactions.  The reaction rates for
these processes  (eq.\ref{11a},\ref{11b})  can  be  used  in  the
kinetic approximation to obtain the shear viscosity coefficients.
In the kinetic approximation,

\begin{equation}  \eta  \approx  \frac{4}{15} n_i <p_i> \lambda_i
\label{15} \end{equation}

\noindent  where  $n_i$  and  $<p_i>$ are the density and average
momentum of the particles of type $i$ and $\lambda_i$ is its mean
free path. Using $<p>=3.2T$  the  viscosity  coefficient  can  be
written as,

\begin{equation}       \eta=       \frac{12.8}{30}      \frac{a_1
T^3}{\frac{R_3}{T}+\frac{R_2}{T}} \label{16} \end{equation}

In  the  following  calculations  we  will use both the viscosity
coefficients eq.\ref{14} and eq.\ref{16}, and will be referred as
vicosity coefficient I and II respectively.

\subsection{Results for RHIC and LHC energies}

The  initial conditions for the hydrodynamic evolution are listed
in table (1) \cite{bi93}. They are the results  of  HIJING  model
calculation,  which is a QCD motivated phenomenological model, as
only initial direct parton scatterings are  taken  into  account.
Thus  there  are some uncertainties in these parameters. However,
they suffice our purpose of demonstrating the effect of viscosity
on parton equilibration process.

In   fig.1a,b   and  c,  we  have  shown  the  evolution  of  the
temperature, the gluon fugacity and the quark  fugacity  at  RHIC
energy.  The  solid  line  corresponds to the ideal flow, i.e. no
viscous drag, the dashed and  the  long  dashed  lines  are  with
viscous drag, corresponding to shear viscosity coefficients I and
II  As  expected,  with  viscosity,  temperature  of the partonic
system  evolve  more  slowly.  Viscosity  oppose  the  expansion,
consequently,  cooling  is  slowed  down. The lifetime of the QGP
phase is considerably increased  in  presence  of  viscosity  (by
nearly  a  factor  of  2).  It  is  apparent  that this will have
considerable effect on signals  of  QGP  such  as  dileptons  and
photons.  We  also  note that temperature evolution do not differ
much with different prescriptions for  $\eta$.  Life  time  being
changed by less than 2\%.

The  evolution  of  gluon fugacity is much more interesting. With
viscosity, gluon fugacity also  evolves  more  slowly.  Viscosity
impedes  the  chemical equilibration process. We find that upto 2
fm, $\lambda_g$ donot change, it then increases slowly, much more
slowly than  for  the  ideal  fluid.  This  is  contrary  to  the
expectation  from  ideal  fluid  flow  (eq.\ref{6})  that  if the
temperature evolve slowly, fugacities will increase at  a  faster
rate.  Indeed  for  viscous  fluid  flow, eq.\ref{6} is no longer
valid. The behavior of $\lambda_g$ can be understood by recasting
eq.\ref{4} as,

\begin{equation}                           \frac{\dot{\lambda}_g}
{\lambda_g}=-3\frac{\dot{T}}{T}-\frac{1}{\tau}               +R_3
(1-\lambda_g)-2R_2                             (1-\frac{\lambda_q
\lambda_{\bar{q}}}{\lambda_g^2}) \label{17} \end{equation}

Then as the temperature decreases more slowly, rhs of eq.\ref{17}
will  decrease  as  a  result of which, fugacity will also evolve
slowly. Thus viscosity, affect both the temperature and  fugacity
evolution,  requiring  them to evolve more slowly than their {\em
ideal} counterpart. We also note that  in  the  viscous  partonic
system,  the gluon fugacity could not attain the value reached in
the ideal case, by the time the system is cooled to  $T_c$,  even
though the life time is considerably increased.

In  fig.1c,  the  quark  fugacity  is  shown. They show a similar
behavior as gluon  fugacity.  As  with  gluon  fugacities,  quark
fugacities  also  evolve  much  more  slowly  if the viscosity is
turned on, for the same reasons as described above. By  the  time
$T_c$  is reached, they are far behind chemical equilibrium. Here
also, the fugacities do not attain the value reached in the ideal
case at $T_c$.

The  results  for  evolution of temperature and fugacities at LHC
energies are nearly the same, as shown in fig.2a,b  and  c.  Here
also,  with  viscosity, plasma cools slowly than the ideal fluid,
life time of the plasma increasing by a factor of  2.  The  gluon
and  quark  fugacities  also  evolve  slowly. In contrast to RHIC
energy, gluon and quark fugacities attain larger values at $T_c$.
This is due to their comparatively large initial values. Thus our
initial conjecture that chemical equilibrium is attained  at  LHC
energies  but not at RHIC is substantiated. However, we note that
even at LHC energies,  the  plasma  is  not  fully  equilibrated,
$\lambda_g \approx 0.7$ and $\lambda_q \approx .5$, and are quite
far from equilibrium.

\section{Thermal photons and lepton pairs}

\subsection{Photon spectra}

Thermal   photons   and   lepton  pairs  are  primary  probes  of
quark-gluon plasma. Being  weakly  interacting,  they  carry  the
information of the early phase of the hot fireball created in the
collision.  Thermal  photons  from  QGP  has  their origin in the
Compton  ($qg\rightarrow  q\gamma$)  and  annihilation  processes
($q\bar{q}\rightarrow  g\gamma$)  processes. For plasma away from
equilibrium, their rates have been calculated \cite{st94}

\begin{equation}
E\frac{dN}{d^3pd^4x}=\frac{2\alpha\alpha_s}{\pi^4}\lambda_q\lambda_g
T^2 e_q^2 exp(-E/T) [\ln(\frac{4ET}{k_c^2})+1/2-C] \end{equation}

\noindent and the rate for annihilation process is,

\begin{equation}
E\frac{dN}{d^3pd^4x}=\frac{2\alpha\alpha_s}{\pi^4}
\lambda_q\lambda_{\bar{q}}        T^2       e_q^2       exp(-E/T)
[\ln(\frac{4ET}{k_c^2})-1-C] \end{equation}

Here  C=0.577721,  $e_q$  is the electric charge of the quark and
the parameter $k_c$ is related to thermal mass of  the  quark  in
the medium,

\begin{equation}           k_c^2=\frac{1}{3}g^2\kappa^2T^2=2m_q^2
\end{equation}

\noindent with  \cite{bi93}

\begin{equation}
m_q^2=(\lambda_g+\frac{1}{2}\lambda_q)\frac{4\pi}{9}\alpha_sT^2
\end{equation}

In  fig.4  and  5,  we have shown the photon spectra, as obtained
presently,  with  and  without  viscous  drag  at  RHIC  and  LHC
energies. It can be seen that photon production is increased by a
factor  of  2 with viscosity turned on. With viscosity turned on,
life time of the plasma is increased nearly by a factor of 2  and
it  is natural that photon yield will also be increased. However,
we have checked that the increase in  photon  yield  beyond  $p_T
\geq$  2  GeV  is  not  due to increased life time of the plasma,
rather, it is due to the slower cooling and  lower  equilibration
rate  of  the  of  the  viscous plasma. The increase in low $p_T$
yield can be attributed to the longer  lifetime  of  the  plasma.
Also we note that, the photon yield do not depend much on the two
version  of  viscosity coefficients used. The yield is marginally
increased at low $p_T$ with viscosity coefficient II,  than  with
viscosity  coefficient  I,  and  essentially  due  to  the larger
lifetime of the plasma.

\subsection{Lepton pair spectra}

The   transverse   mass   distribution   of   lepton  pairs  from
nonequilibrium plasma can be written as \cite{st94},

\begin{equation}  \frac{dN}{dM^2d^2M_Tdy}=\frac{\alpha^2}{4\pi^3}
e_q^2  R_T^2  \int   \lambda_q   \lambda_{\bar{q}}   \tau   d\tau
K_0(\frac{M_T}{T}) \end{equation}

In  fig.5 and 6, we have shown the invariant mass distribution of
lepton pairs at RHIC and LHC energies. As before, we  have  shown
the  results  with  and without viscous drag. Here again, we find
that the lepton pair yield in increased by a factor  of  2,  when
the  viscosity  is  turned  on. As found with the photon spectra,
here also, large mass ($M \geq$ 2 GeV) yield of of  lepton  pairs
are  essentially  due to slower cooling and equilibration rate of
the viscous plasma,  rather  than  its  increased  lifetime.  The
increase  in  the  low  mass  yield  can be said to be due to the
increased lifetime of the plasma. As  with  photon  spectra,  the
lepton  pair  yield  with two different viscosity coefficients do
not differ much.

The  analysis  clearly indicate that photon and dilepton yield in
QGP are greatly affected if dissipative effects like viscosity is
taken into account. Large invariant mass lepton  pairs  and  also
large $p_T$ photons yields are increased by a factor of two, from
their  ideal  counterpart.  This  increase  is essentially due to
lower cooling and equilibration rate. The low $p_T$  photons  and
the  low  invariant mass dileptons are also increased, but mainly
due to increased life time of the viscous plasma.

\section{Summary}

We   have  investigated  the  effect  of  viscosity  on  chemical
equilibration of parton distribution in  relativistic  heavy  ion
collisions.  We  assumed  that  after  a  time  $\tau_{iso}$, the
partonic system achieved kinetic equilibrium. Beyond $\tau_{iso}$
expansion (which we assumed to  be  boost-invariant  longitudinal
only)  is governed by the hydrodynamic equations. The approach to
the chemical equilibration process is then governed by a  set  of
master  equations  which  are  driven  by  the two body reactions
$gg\leftrightarrow q\bar{q}$  and  $gg\leftrightarrow  ggg$.  The
partonic  matter continue to expand and cool due to expansion and
chemical equilibration, till the critical temperature  ($T_c$=160
MeV)  is  reached. Initial condition of the plasma was taken from
HIJING calculations. It was seen that with viscosity  turned  on,
the   temperature  of  the  partonic  system  evolve  slowly,  in
comparison to the ideal system. This is  expected,  as  viscosity
impedes  the  expansion,  making  the  cooling  a slower process.
Interestingly, we find that the chemical  equilibration  rate  is
also slowed down in presence of viscous drag. This is in contrast
to  the ideal partonic system behavior, where, lower cooling rate
implies a faster equilibration. We thus find that even though the
life time of the plasma is increased by nearly a factor of 2, the
gluon or quark fugacity  could  not  attain  the  value  that  is
reached in case of ideal flow.

The  effect of lower cooling and equilibration rate on photon and
lepton pair production was also studied. It was  seen  that  with
viscosity,  the  photon as well as lepton pair yield increased by
nearly a factor of 2, both at RHIC and LHC energies. The increase
in large $p_T$ photons or  large  invariant  mass  dileptons  are
essentially  due to the lower cooling rate, rather than the large
life time of the viscous plasma. The increase in  the  low  $p_T$
photon  or  low invariant mass dileptons can be attributed to the
to the longer life of the plasma.

To  conclude,  viscosity  can have a significant effect on parton
equilibration process, by slowing down the cooling rate  and  the
chemical  equilibration  process.  Its effect is also manifest in
the pre-equilibrium photon and dilepton production, raising their
yield by nearly a factor of 2.

\begin{table}  \text{Initial conditions characterising the parton
plasma    at    the    onset    of    hydrodynamic     evolution}
\begin{tabular}{ccc}  &RHIC&LHC\\ \tableline $\tau_{iso}$(fm/c) &
0.31 & 0.23\\ $T_0$(GeV) & 0.57 & 0.83\\  $\lambda_g^0$&  0.09  &
0.14\\ $\lambda_q^0$ & 0.02 & 0.03 \end{tabular} \end{table}
\begin{figure}
\caption{Evolution of temperature, gluon fugacity and quark
fugaciy with proper time at RHIC.}
\end{figure}
\begin{figure}
\caption{Evolution of temperature, gluon fugacity and quark
fugaciy with proper time at LHC.}
\end{figure}
\begin{figure}
\caption{Transverse momentum distribution of photons at RHIC}
\end{figure}
\begin{figure}
\caption{Transverse momentum distribution of photons at LHC}
\end{figure}
\begin{figure}
\caption{Invariant mass distributions of lepton pairs at RHIC.}
\end{figure}
\begin{figure}
\caption{Invariant mass distributions of lepton pairs at LHC.}
\end{figure}
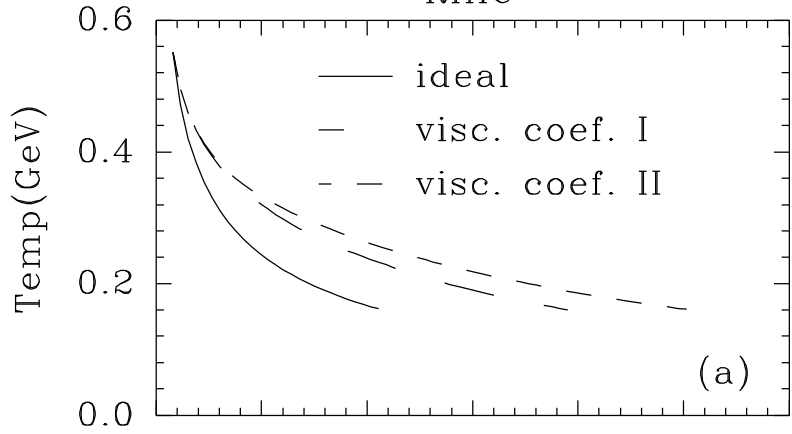
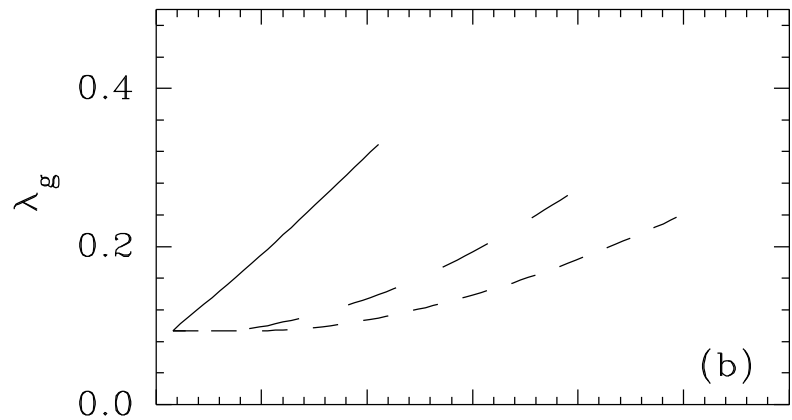
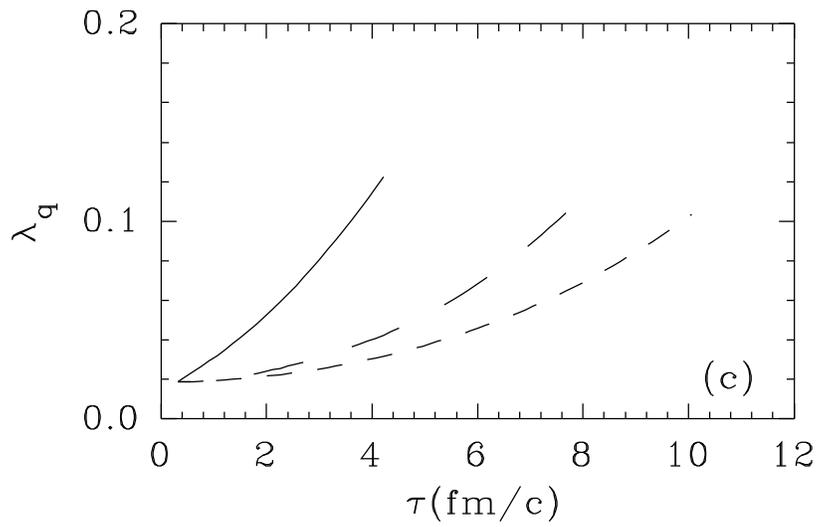
\end{document}